\documentclass[prl,superscriptaddress,twocolumn]{revtex4-1}

\usepackage{graphicx}
\usepackage{mathtools}
\usepackage{color}
\usepackage{amsmath}
\def\vector#1{\mbox{\boldmath $#1$}}

\begin{document}
\title{Speedup of nuclear-spin diffusion in hyperpolarized solids}
\author{Yu Wang}
\author{Kazuyuki Takeda}
\email{takezo@kuchem.kyoto-u.ac.jp}
\affiliation{Division of Chemistry, Graduate School of Science, Kyoto University, 606-8502 Kyoto, Japan}
\date{\today}
\begin{abstract}
We propose a correction to the coefficient of nuclear-spin diffusion by a factor $1/\sqrt{1-\overline{p}^{2}}$, where $\overline{p}$ is the average nuclear-spin polarization. The correction, derived by extending the Lowe-Gade theory to low-temperature cases, implies that transportation of nuclear magnetization through nuclear-spin diffusion accelerates when the system is hyperpolarized, whereas for low-polarization the correction factor approaches unity and the diffusion coefficient coincides with the conventional diffusion coefficient valid in the high-temperature limit. The proposed scaling of the nuclear-spin diffusion coefficient can lead to observable effects in the buildup of nuclear polarization by dynamic nuclear polarization.
\end{abstract}

\maketitle

Nuclear-spin diffusion\cite{Bloembergen1949} plays an essential role in nuclear-magnetization transfer, nuclear-spin relaxation, and dynamic nuclear polarization (DNP) in rigid solids composed of spin-active isotopes, where nuclear spins interact magnetically with one another through the dipole-dipole couplings.
In the presence of an externally applied magnetic field, the effect of the dipolar interaction between a pair of adjacent nuclear spins precessing at a common Larmor frequency is to cause exchange of their spin states.
In a time scale longer than that of the mutual spin flips, exchanges of spin-states among a number of nuclei lead to diffusive transportation of the nuclear-spin polarization and thereby of the longitudinal nuclear magnetization without accompanying mass motion.

When position-dependent nuclear-spin polarization $P(\mathbf{r})$ develops for some reasons, e.g., by selective saturation in heterogeneous systems like polymer blends, creation of polarization-grating\cite{Zhang1998}, or DNP in solids, 
nuclear-spin diffusion, governed by the diffusion equation
\begin{align}
    \frac{\partial}{\partial t} P = D \Delta P,
    \label{eq:diffusion-eq}
\end{align}
tends to flatten the inhomogeneous profile of nuclear-spin polarization toward uniform one.
Here, $D$ is the spin-diffusion coefficient.
Since nuclear-spin diffusion is driven by the dipole-dipole interactions, and since the dipolar interaction is through-space, i.e., characterized only by the magnitude and orientation of the internuclear vector between the relevant spins, the nuclear spin-diffusion coefficient has so far been thought of being determined solely by the crystal structure.
Indeed, formulas for the diffusion coefficient using the coordinates of the nuclear spins involved in the system of interest were successfully derived under the high-temperature approximation\cite{Lowe1967,Redfield1968}.
Here, we point out that $D$ needs to be modified at low nuclear-spin temperature, or equivalently, when the nuclear spin system is hyperpolarized.
As we discuss below, $D$ would be affected by the average nuclear polarization $\overline{p}$ in such a way that it is scaled by a factor $1/ \sqrt{1-\overline{p}^{2}}$, so that $D$ is represented as
\begin{align}
    D = \frac{D_{0}}{\sqrt{1 -  \overline{p}^{2}}},
    \label{eq:DC}
\end{align}
where $D_{0}$ is the conventional diffusion coefficient valid in the high-temperature limit, and $D$ and $D_{0}$ coincide when the nuclear polarization is low $(\overline{p} \ll 1)$.

The implication of the correction factor $1/\sqrt{1-\overline{p}^{2}}$ for hyperpolarized nuclei would be such that the spin-diffusion coefficient, often referred to as the spin-diffusion \textit{constant}, is \textit{no longer} a constant.
As increasing the overall nuclear polarization, $D$ would also be increased, i.e., spatial transportation of nuclear magnetization by nuclear-spin diffusion would become faster.
The way that $D$ increases with $\overline{p}$ would become more prominent as the latter approaches unity.
Even though $D$ in Eq.~(\ref{eq:DC}) becomes infinity with $\overline{p}=1$, the singularity would not be a problem, because then the polarization would be perfectly uniform, making spin diffusion irrelevant.

Let us consider a rigid system composed of identical spins $I=\frac{1}{2}$ in a static magnetic field, and represent the position of the $i$-th spin by $\vector{r}_{i} = (x_{i}, y_{i}, z_{i})$, the relative position between the $i$-th and $j$-th spins by $\vector{r}_{ij}=(x_{j}-x_{i}, y_{j}-y_{i}, z_{j}-z_{i})$, and the angle of $\vector{r}_{ij}$ with respect to the static field by $\theta_{ij}$.
Index $i=0$ is reserved for the site chosen as the origin.
According to Lowe and Gade\cite{Lowe1967,Lowe1968,Lowe1972}, the $(\alpha, \beta)$-component ~$(\alpha,\beta=x,y,z)$ of the spin diffusion coefficient $(D_{0})_{\alpha,\beta}$ in the high-temperature limit is given, up to the second order, by
\begin{align}
  (D_{0})_{\alpha,\beta} = \frac{\sqrt{\pi}}{2}
                     \sum_{i \neq 0} A_{i} \alpha_{i} \beta_{i} F_{i},   \label{eq:DLG}
\end{align}
with
\begin{align}
  A_{i} &= A_{0,i}, \\
  A_{ij} &= \left( \frac{\mu_{0}}{4\pi} \right) \left(-\frac{1}{4}\right) \gamma^{2} \hbar r_{ij}^{-3} \left( 1 - 3\cos^{2}\theta_{ij} \right). \label{eq:A}
\end{align}
Here, $\gamma$ is the gyromagnetic ratio, and $\mu_{0}$ is the vacuum permeability.
The expression for $F_{i}$ in Eq.~(\ref{eq:DLG}), given by
\begin{align}
  F_{i} &\approx \frac{\sqrt{\pi}}{2} \left[ \frac{1}{2} \sum_{j \neq i} \left( B_{j} - B_{ij} \right)^{2} \right]^{-\frac{1}{2}}, \label{eq:F} \\
  B_{ij} &= -2 A_{ij}, ~ B_{j} = B_{0,j},
\end{align}
comes from
\begin{align}
    F_{i} &= \int_{0}^{t} \langle G_{i}(\tau) \rangle d\tau, \label{eq:F2} \\
    G_{i}(\tau) &= \mathrm{Re}
      \{
        \prod_{j \neq i}
        \left[
          \cos(B_{j}-B_{ij})\tau  \right. \nonumber \\
           & \ \left. + 2i \langle I_{jz} (t-\tau) \rangle \sin(B_{j}-B_{ij})\tau
        \right]
      \} \label{eq:G},
\end{align}
where, in the limit of low spin polarization $p_{j} \equiv 2 \langle I_{jz} \rangle \ll 1$, the sine terms in Eq.~(\ref{eq:G}) are negligible.
The profile of $G_{i}(\tau)$, which is unity at $\tau=0$, is such that it decays to values much smaller than 1 by interference between the multiple cosine terms, and eventually, recovers to unity again at $\tau_{\mathrm{p}}=\prod_{j\neq 0,i} 2\pi/(B_{j}-B_{ij})$.
The period $\tau_{\mathrm{p}}$ of recovery increases with the size of the system, and is expected to become exceedingly long compared to the decoherence time as the number of the spins involved in the system increases.
From a physical point of view, it would therefore be fairly reasonable to focus on the spin dynamics over a time interval much shorter than $\tau_{\mathrm{p}}$ and discard the dynamics afterwards, to approximate $G_{i}(\tau)$ by a decaying function that \textit{never recovers}, and to take the integral in Eq.~(\ref{eq:F2}) up to infinity, i.e., $t \rightarrow \infty$.
Retaining the cosine terms in Eq.~(\ref{eq:G}), we expand $G_{i}$ as $1 - \frac{1}{2} \sum_{j \neq 0,i} (B_{j} - B_{ij})^{2} \tau^{2} + O(\tau^{4})$.
Such a decaying profile is well reproduced by introducing a Gaussian function $G_{i} \approx \exp(-\Delta_{i} \tau^{2})$, if we request $\Delta_{i}$ to be
\begin{align}
    \Delta_{i} = \frac{1}{2} \sum_{j \neq 0,i} (B_{j} - B_{ij})^{2},
\end{align}
whence, from Eq.~(\ref{eq:F2}), Eq.~(\ref{eq:F}) is obtained.


We now turn our attention to a hyperpolarized spin system, in which $p_{j}=2\langle I_{jz} \rangle$ in Eq.~(\ref{eq:G}) can no longer be neglected.
We represent the local polarization $p_{j}$ at the $j$-th site as $\overline{p} + \delta p_{j}$, where $\overline{p}$ is the average polarization common to all spins, while $\delta p_{j} \ll 1$ is the local deviation at the $j$-th site.
Then, we can rewrite $G_{i}(\tau)$ as (see Supplemental Material)
\begin{align}
    G_{i}(\tau) &\approx
    \left[
        \prod_{j \neq i}
        \cos(B_{j}-B_{ij})\tau
    \right] \nonumber \\
    & \cdot
    \left[
        1 + \overline{p}^{2} \frac{1}{2}
        \sum_{l \neq i} \tan^{2}(B_{l}-B_{il})\tau
    \right].
    \label{eq:G3}
\end{align}
We note that the first term in Eq.~(\ref{eq:G3}) coincides with $G_{i}(\tau)$ in the high-temperature approximation, while the second term can now be approximated by $\exp(\overline{p}^{2}\Delta_{i}\tau^2)$, so that
\begin{align}
  G_{i}(\tau) \approx \exp\left[-(1-\overline{p}^{2}) \Delta_{i} \tau^{2} \right].
\end{align}
Then, we obtain $F_{i}$ to be $\frac{\sqrt{\pi}}{2} [(1-\overline{p}^{2}) \Delta_{i}]^{-1/2}$, and therefore the correction factor of the spin-diffusion coefficient to be $1/\sqrt{1 - \overline{p}^{2}}$, as proposed in Eq.~(\ref{eq:DC}).

Another formula for the spin-diffusion coefficient in the high-temperature limit, put forth by Redfield, is based on the response of a bulk spin system to a sinusoidally space-time varying field\cite{Redfield1959,Redfield1968}.
For spin-$\frac{1}{2}$ nuclei, the high-temperature spin-diffusion coefficient $D_{0}$, given by
\begin{align}
    D_{0} = \frac{3\sqrt{\pi}\sum_{j} x_{ij}^{2} A_{ij}^{2}}{16 \sqrt{\langle \Delta \omega ^{2}\rangle}},
    \label{eq:DC-R}
\end{align}
include the square root of the Van Vleck second moment $\langle \Delta \omega^{2}\rangle$\cite{VanVleck1948} in the denominator, and gives similar values to those calculated by the Lowe-Gade formula.

The Van Vleck second moment measures line broadening of nuclear magnetic resonance (NMR) spectra of homonuclear spin systems.
Interestingly, polarization dependence of the resonance line, obtained either by continuous-wave NMR or by pulsed NMR using a small-tip pulse\cite{Lowe1957,Waugh1987}, is such that the dipolar linewidth as well as the second moment $\langle \Delta \omega^{2}\rangle$ scales with the average polarization $\overline{p}$ as $1-\overline{p}^{2}$, as proved by Abragam et al., who also generalized the formula for the second moment\cite{Abragam1973}.
Thus, one may be tempted to naively replace the Van Vleck second moment in Eq.~(\ref{eq:DC-R}) by the Abragam's version, which would then give the same correction factor that we propose above.
However, no reasonable justification for such replacement can be found, because the diffusion coefficient $D_{0}$ in Eq.~(\ref{eq:DC-R}) is the outcome based on the high-temperature approximation, so that the very factors that become significant when the system is polarized have already been discarded in the process of deriving Eq.~(\ref{eq:DC-R}).
Nevertheless, such a coincidence implies that successful extension of the Redfield theory in future to hyperpolarized systems should be accompanied by the Abragam's second moment.

\begin{figure*} [htbp]
\begin{center}
\includegraphics[width=0.95\linewidth]{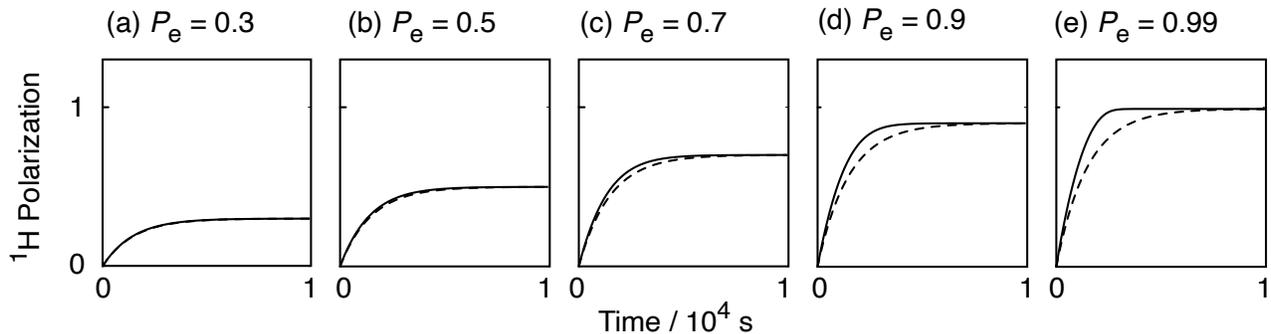}
\caption{Simulated buildup curves of nuclear polarization in a periodic box with side length $L=2\times10^{-8}~\mathrm{m}$, nuclear-spin density $\rho=4.36\times10^{26}~\mathrm{m}^{-3}$, and the high-temperature spin-diffusion coefficient $D_{0}=4.5\times10^{-19}~\mathrm{m}^{2}\mathrm{s}^{-1}$. Inside the box a single source passes its polarization, $P_{\mathrm{e}}$, to the nuclear spin at its nearest neighbour. The source polarization $P_{\mathrm{e}}$ was (a) 0.3, (b) 0.5, (c) 0.7, (d) 0.9, and (e) 0.99. Solid and broken lines are results obtained with and without the scaling of the nuclear-spin diffusion coefficient according to Eq.~(\ref{eq:DC}).
}
\label{fig:sim1}
\end{center}
\end{figure*}

In order to show that the polarization-dependent nuclear-spin diffusion rate can indeed lead to the outcomes that can be tested experimentally, we focus on DNP in solids containing a small fraction of paramagnetic electrons.
Even though only a limited number of nuclei being relatively close to the paramagnetic electrons can acquire the electron-spin polarization directly, nuclear-spin diffusion carries such locally created high polarization to the more distant nuclei.
It is the combination of the direct electron-to-nucleus polarization transfer and nuclear-spin diffusion that eventually leads to bulk nuclear hyperpolarization.
The buildup behavior of nuclear polarization and thereby that of the intensity of the NMR signal ought to be quantitatively explained by taking account of both of these two processes as well as spin-lattice relaxation.

We consider a model in which a cubic region with side length $L$ contains $\rho L^{3}$ nuclear spins, where $\rho$ is the density of the latter, and there exists a single paramagnetic electron spin in the box, serving for the source of polarization.
We assume that the nuclear-spin polarization is initially zero, and the electron-spin polarization $P_{\mathrm{e}}$ can be arbitrarily set by adjusting the temperature and/or the magnetic field strength.
The channel of polarization transfer from the electron spin to the nuclear spin at the nearest neighbour of the former can be connected either continuously or intermittently by switching on/off microwave irradiation in such a way that the solid effect, thermal mixing effect, cross effect, or integrated solid effect drives electron-to-nucleus polarization transfer with a certain efficiency.
Meanwhile, the nuclear-spin polarization evolves in time according to the spin-diffusion equation Eq.~(\ref{eq:diffusion-eq}) under the periodic boundary condition.
By numerical simulations, we examine the buildup behavior of the overall nuclear-spin polarization with/without the correction to the spin-diffusion coefficient according to Eq.~(\ref{eq:DC}).

Figure~\ref{fig:sim1} shows source-polarization dependence of simulated buildup curves based on this model with a set of parameters that mimics a $^{1}$H spin system in partially deuterated solid material containing a fraction of paramagnetic impurities\cite{Pinon2020}.
Here, spin-lattice relaxation was ignored.
As a result, the nuclear polarization is eventually built up to the ultimate value given by the electron source polarization $P_{\mathrm{e}}$ for all cases.
For relatively low source polarization, the profiles of the simulated buildup curves with/without correction to the spin-diffusion coefficient were similar.
Indeed, for $P_{\mathrm{e}}=0.3$, the two buildup curves overlapped, as shown in Fig.~\ref{fig:sim1}(a).
As increasing the source polarization, the buildup curves simulated with the corrected spin-diffusion coefficient began to show slight deviation from those obtained without correction, exhibiting faster buildup compared to the latter (Fig.~\ref{fig:sim1}(b)).
Such acceleration of the buildup behavior became more prominent as the source polarization $P_{\mathrm{e}}$ further increased, as demonstrated in Fig.~\ref{fig:sim1}(c)-(e).
For $P_{\mathrm{e}}=0.5$, the spin-diffusion coefficient would become ca.~1.15 times larger at the final stage of the buildup compared to that would be at the initial stage when the nuclear polarization is much lower than 1.
The scaling factor for $P_{\mathrm{e}}=0.7, 0.9$, and $0.99$ would be 1.4, 2.3, and 7.1.

When one aims to characterize DNP buildup curves in practice, one is required to take spin-lattice relaxation into account.
In addition, the rate $\xi$ of the local electron-to-nucleus polarization transfer and the high-temperature spin-diffusion coefficient $D_{0}$ need to be determined.
There three parameters can be obtained from separate measurements.
For the spin-lattice relaxation rate, the standard saturation/inversion recovery methods can be used.
The exchange rate $\xi$ can be determined from DNP experiments employing intermittent applications of microwave pulses, so that the period of the electron-to-nucleus polarization transfer and that of nuclear-spin diffusion follow one after another\cite{Miyanishi2021}.
Given the knowledge of the exchange rate $\xi$, examination of the initial buildup rates for various repetition rates of pulsed microwave applications allows for determination of the high-temperature diffusion coefficient $D_{0}$\cite{Kagawa2009,Miyanishi2021}.
Creation of magnetization grating is another way for experimentally determining the high-temperature spin-diffusion coefficient\cite{Zhang1998}.

The exchange rate $\xi$, the high-temperature spin diffusion coefficient $D_{0}$, and the spin-lattice relaxation rate, are sufficient for simulating the buildup behavior of nuclear polarization.
Alternatively, given two of these three parameters, the rest can be used as an adjustable parameter in the simulations so as to reproduce the experimental data obtained under fast repetition of the microwave pulses, or continuous application of microwave irradiation, where the nuclear spin system can be optimally polarized.

\begin{figure} [htbp]
\begin{center}
\includegraphics[width=0.9\linewidth]{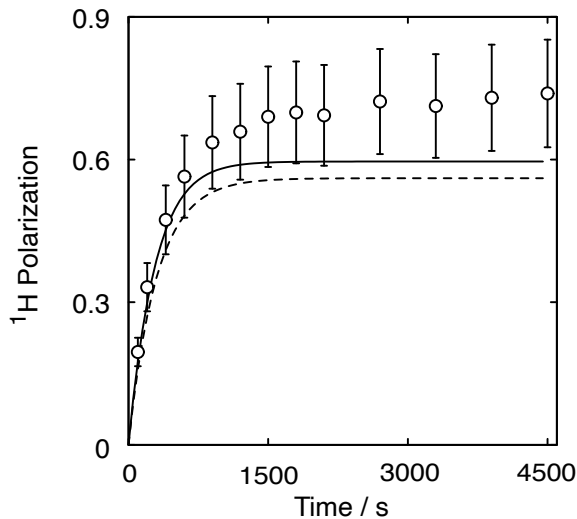}
\caption{Circles represent experimental buildup of $^{1}$H polarization by triplet DNP, reported in Ref.~\cite{Takeda2004b}, in single crystal of 99.21 \% deuterated naphthalene doped with pentacene with concentration of $1.46 \times 10^{-4}$ in 0.3187 T and at 100 K. Solid and broken lines represent simulated buildup curves with and without the proposed scaling of the nuclear-spin diffusion coefficient. Experimental details are described in Supplemental Material.} \label{fig:sim2}
\end{center}
\end{figure}

Figure.~\ref{fig:sim2} shows experimental buildup behavior of $^{1}$H polarization by DNP reported in Ref.~\cite{Takeda2004b}, where the residual $^{1}$H spins in 99.21 \% deuterated single crystal naphthalene, doped with pentacene, were polarized by DNP.
In this what is called triplet DNP experiment, the electron spins in the photo-excited triplet state of pentacene served for the source of polarization, which is as high as ca.~0.7.
During the first 15~$\mu$s of the lifetime (ca.~100 $\mu$s) of the photo-excited triplet state of pentacene, a microwave pulse was applied together with magnetic-field sweep to induce the integrated solid effect (ISE)\cite{Henstra1988,Henstra1990}.
The cycle of pulsed laser excitation, ISE, and spin diffusion was repeated at a rate of 50 Hz, and the resultant $^{1}$H polarization was recorded as a function of the buildup time.
The considerable range of the error stems from slight uncertainty in the size $2.7 \pm 0.2$ mm of the diameter of the laser beam applied at the crystal for photo-excitation, which affected the active volume of the sample and thereby the enhancement factor of the NMR signal intensity compared to that measured for the dark sample in thermal equilibrium.

In Fig.~\ref{fig:sim2} also plotted are simulated curves based on the model described above, with and without correction to the diffusion coefficient.
In this system, the spin-lattice relaxation rate $1/T_{1}$ and the high-temperature $^{1}$H spin-diffusion coefficient $D_{0}$ were found to be $7.6 \times 10^{-4}$~s$^{-1}$ and $4.5 \times 10^{-19}$~m$^{2}$s$^{-1}$, respectively (See Supplemental Material).
The exchange probability $\xi$ was set to 1, which was found to reproduce the initial slope of the experimental buildup curve.
The profile of the simulated buildup curve with the scaled spin-diffusion coefficient shows higher final polarization compared to that obtained with the conventional, high-temperature spin-diffusion coefficient.
Here, the finite spin-lattice relaxation tends to drag the polarized nuclear-spin system back toward thermal equilibrium.
If the spin-diffusion rate increases with polarization, the spin system would gain more chance to distribute the hyperpolarization.
Therefore, such a nuclear-hyperpolarization experiment can serve for a test of whether the proposed scaling of the nuclear-spin diffusion coefficient according to Eq.~(\ref{eq:DC}) is legitimate, provided that the parameters governing the buildup behavior are well characterized, and that the nuclear-spin polarization is accurately determined.
Unfortunately, the considerable experimental error in the data in Fig.~\ref{fig:sim2} does not allow for convincing verification of our theory.
Nevertheless, the current work poses a timely open question, now that DNP has become widespread and quite a few groups are equipped with such facilities that could implement nuclear hyperpolarization experiments\cite{Pinon2020}.

To summarize, the hitherto belief that the nuclear spin-diffusion coefficient is determined solely by the coordinates of the relevant nuclei needs to be modified when the spin system is hyperpolarized.
The proposed correction factor $1/\sqrt{1-\overline{p}^2}$, derived by the low-temperature extension of the Lowe-Gade theory, albeit not being appreciable for $\overline{p} \ll 1$, would become significant as $\overline{p}$ approaches unity.
The correction implies that the rate of transportation of nuclear magnetization by spin diffusion in rigid lattice ought to increase with the overall average spin polarization.
Buildup of nuclear polarization by DNP is an example of the observable effects of the scaling of the nuclear-spin diffusion coefficients.

\begin{acknowledgments}
This work has been supported by MEXT Quantum Leap Flagship Program (MEXT Q-LEAP) (Grant Number JPMXS0120330644) and JST CREST (Grant Number JPMJCR1873).
\end{acknowledgments}









\hrulefill
\clearpage
\clearpage
\newpage
\widetext

\renewcommand{\thefigure}{S\arabic{figure}}
\setcounter{figure}{0}

\renewcommand{\theequation}{S\arabic{equation}}
\setcounter{equation}{0}

\widetext


\begin{center}
\begin{LARGE}
\textbf{Supplemental Material: Speedup of nuclear-spin diffusion in hyperpolarized solids}
\end{LARGE}
\end{center}

\section{Derivation of Eq.~(11)}
Let us recall Eq.~(9) for the expression of $G_{i}(\tau)=\mathrm{Re}\{ H_{i}(\tau) \}$, with $H_{i}(\tau)$ being
\begin{align}
  H_{i}(\tau) =
    \prod_{j \neq i}
        \left[
          \cos(B_{j}-B_{ij})\tau
         + 2i \langle I_{jz} \rangle \sin(B_{j}-B_{ij})\tau
        \right],
  \label{eq:H}
\end{align}
or,
\begin{align}
  H_{i}(\tau) &=
   \left[ \cos(B_{1}-B_{i,1})\tau + 2 i \langle I_{1z} \rangle \sin(B_{1}-B_{i,1})\tau \right] \nonumber \\
   &\times \left[ \cos(B_{2}-B_{i,2})\tau + 2 i \langle I_{2z} \rangle \sin(B_{2}-B_{i,2})\tau \right] \nonumber \\
   & \cdots \nonumber \\
   &\times \left[ \cos(B_{i-1}-B_{i,i-1})\tau + 2 i \langle I_{i-1,z} \rangle \sin(B_{i-1}-B_{i,i-1})\tau \right] \nonumber \\
   &\times \left[ \cos(B_{i+1}-B_{i,i+1})\tau + 2 i \langle I_{i+1,z} \rangle \sin(B_{i+1}-B_{i,i+1})\tau \right] \nonumber \\
   & \cdots
\end{align}
In the original work by Lowe and Gade, the high-temperature approximation was made and $\langle I_{jz} \rangle = \frac{1}{2} p_{j}$ was assumed to be negligible.
Here, we represent the local polarization $p_{j}$ at site $j$ as $\overline{p} + \delta p_{j}$, where $\overline{p}$ is a common polarization which can become as high as being close to 1, and $\delta p_{j} \ll 1$ is the local fluctuation.
We represent $G_{i}(\tau)$ retaining terms of up to the second order with respect to $\overline{p}$.
Noting that all of the first order terms are pure imaginary and therefore have no contribution to $G_{i}(\tau)=\mathrm{Re}\{ H_{i}(\tau) \}$, we obtain
\begin{align}
    G_{i}(\tau) \approx&
    \prod_{j \neq i} \cos(B_{j}-B_{ij})\tau \nonumber \\
    &- \overline{p}^{2}\sum_{l \neq i} \sum_{m \neq i,l} \left[ \sin(B_{l}-B_{il})\tau \sin(B_{m}-B_{im})\tau \prod_{p \neq i,l,m} \cos(B_{p}-B_{ip})\tau \right] \\
    =& \prod_{j \neq i} \cos(B_{j}-B_{ij})\tau \left[ 1 - \overline{p}^{2} \sum_{l \neq i} \tan(B_{l}-B_{il})\tau \sum_{m \neq i,l} \tan(B_{m}-B_{im})\tau \right]. \label{eq:G3-s}
\end{align}

\begin{figure} [htbp]
\begin{center}
\includegraphics[width=0.4\linewidth]{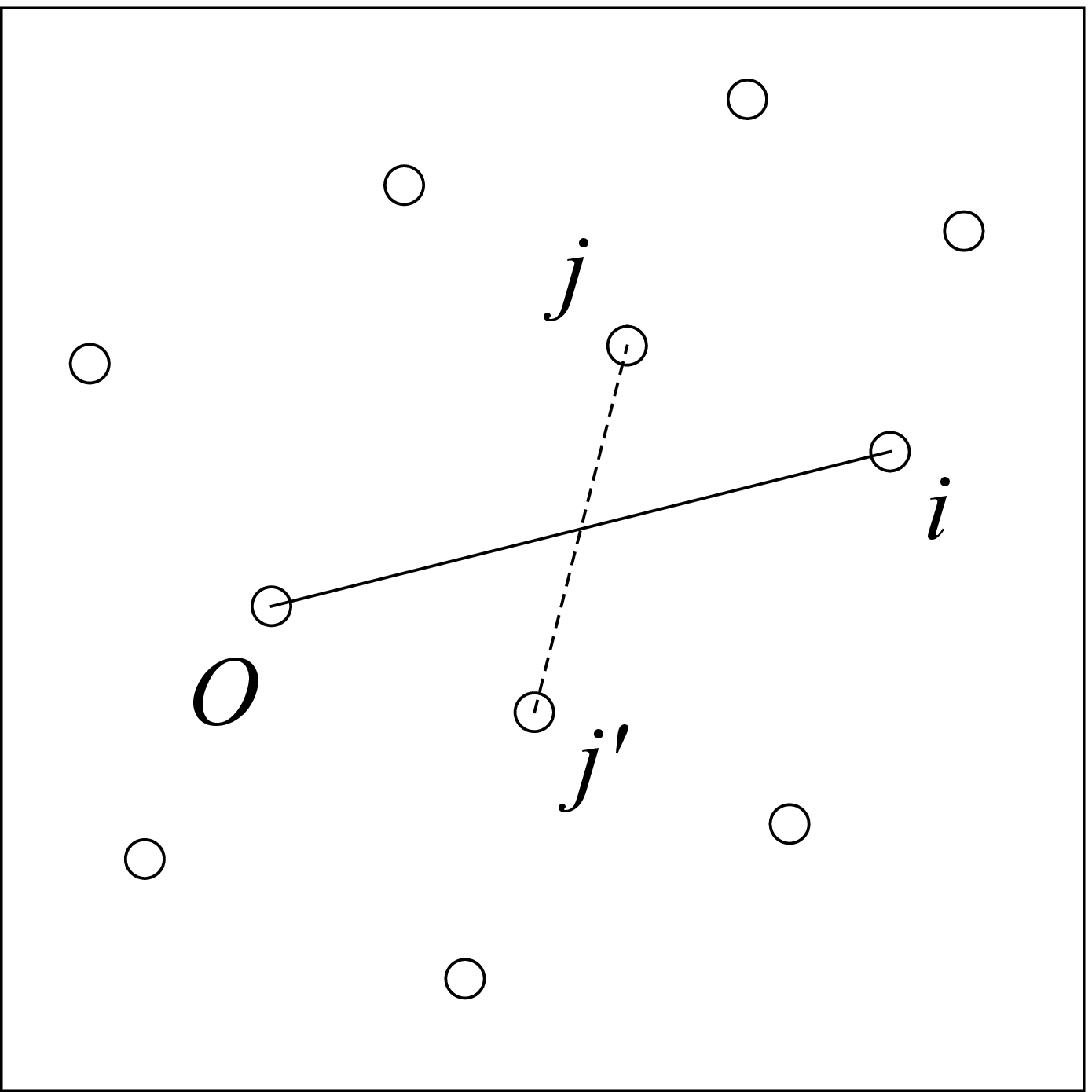}
\caption{Description of the idea of inversion symmetry. Circles correspond to the sites where the individual nuclear spins are located. $O$ represents the origin. The sites, which would be distributed in three-dimensional space in reality, are schematically drawn in a plane for simplicity.}
\label{fig:s1}
\end{center}
\end{figure}

To evaluate $G_{i}$ for a given set of coordinates of the relevant nuclear spins and arbitrarily chosen origin $O$,
let us now suppose that the distribution of the nuclear-spin sites is symmetric and that, for every site $j (\neq O, i)$, there exists another site, $j'$, such that sites $j$ and $j'$ constitute a pair of inversion symmetry with respect to the central point of the line segment connecting $O$ and $i$, as schematically described in Fig.~\ref{fig:s1}.
Then, all terms in the summation over index $m \neq i,l$ in Eq.~(\ref{eq:G3-s}), except for the summand for site $m=l'$ corresponding to the inversion symmetry of site $l$, have their own counterpart such that
\begin{align}
    \tan(\underbrace{B_{m'}}_{=B_{im}}-\underbrace{B_{im'}}_{=B_{m}})\tau = -\tan(B_{m}-B_{im})\tau.
\end{align}
It follow that all but one terms in the sum over index $m$ cleanly cancel and vanish, and we have the only non-vanishing term for $m=l'$, i.e.,
\begin{align}
    \tan(\underbrace{B_{l'}}_{=B_{il'}}-\underbrace{B_{il'}}_{=B_{l}})\tau = -\tan(B_{l}-B_{il})\tau,
\end{align}
leading to
\begin{align}
    G_{i}(\tau) \approx&
      \prod_{j \neq i} \cos(B_{j}-B_{ij})\tau \left[ 1 + \overline{p}^{2} \sum_{l \neq i} \tan^{2}(B_{l}-B_{il})\tau \right]. \label{eq:G4}
\end{align}



\section{Experimental details}
99.21~\% deuterated naphthalene was purified by zone-melting, and pentacene-doped single crystal was grown by the Bridgman method.
The concentration of pentacene was $1.46 \times 10^{-4}$.
The naphthalene crystal, known to be monoclinic, was cut into a piece with a size of $4 \times 4 \times 2.2$~mm, and the $4 \times 4$~mm $ab$-plane showing cleavage was mounted on a uniaxial goniometer such that the $b$-axis, found by inspecting birefringence, was parallel to the goniometer axis.
The crystal was aligned in such a way that the long axis of the pentacene molecules, at both of the two possible substitution sites in the unit cell of naphthalene, are aligned along the external magnetic field, by tilting the goniometer by ca.~23 degrees from the orientation at which the $ab$ plane is normal to the external field.

The guest pentacene molecules were photo-excited by a 590 nm pulsed laser beam with a power of 5 mJ/pulse.
The pulse width of the laser was 1 $\mu$s, and the diameter of the beam on the sample was $2.7\pm0.2$~mm.
Care had been paid to make sure that the penetration depth and \textit{the triplet-excitation depth}, the distance from the sample surface to which the guest molecules are photo-excited and also undergo intersystem crossing to the triplet state, was sufficiently long compared to the sample thickness\cite{s-Takeda2002}.
During the lifetime of the metastable triplet state, microwave irradiation and magnetic-field sweep were applied in such a way that the integrated solid effect (ISE)\cite{Henstra1988,Henstra1990} induces polarization transfer from the electron spins in the photo-excited pentacene to the residual protons with an abundance of 0.79 \%.

Microwave irradiation was applied at the sample using a loop-gap resonator immediately after the laser pulse for 15 $\mu$s.
The frequency of the microwave resonator, 9.70 GHz, determined the required magnetic field to be 0.3187 T, which corresponds to the resonance condition of the electron spins in the photo-excited triplet state of pentacene with its long molecular axis aligned along the static field.
During microwave irradiation, magnetic-field sweep with a width of 7.2 mT was performed.
The cycle of photo-excitation and ISE was repeated at a rate of 50 Hz, and the enhanced $^{1}$H NMR signals were monitored by a pulsed NMR technique.
During the experiment, nitrogen gas cooled at 100 K was brown at the sample.

The $^{1}$H polarization was estimated by comparing the amplitude of the NMR signal per $^{1}$H spin of the sample with that of H$_2$O reference in thermal equilibrium at room temperature.

\section{Validity of the Lowe-Gade theory for systems with two different spin species}
In the sample for which $^{1}$H-polarization buildup experiment by triplet-DNP is demonstrated in Fig.~2, a large number of the $^{2}$H spins surrounding the residual $^{1}$H spins cause $^{1}$H-$^{2}$H heteronuclear dipolar broadening of the $^{1}$H resonance line.
As a consequence, the rate of mutual spin flip-flop among the $^{1}$H spins is decreased, and therefore the $^{1}$H spin diffusion becomes slower than it would be if it had not been for the $^{2}$H spins.
Conversely, $^{1}$H spin diffusion in heavily deuterated organic solids was shown to be accelerated under application of $^{2}$H double-quantum decoupling\cite{s-Negoro2010}.
In this work, we discuss the scaling of the spin-diffusion rate with nuclear-spin polarization based on the Lowe-Gade theory, which originally deals with an assembly of dipolar-coupled homonuclear spins.
Now, a question that may naturally arise is: is the Lowe-Gade theory also applicable to systems composed of two different spin species?

In high magnetic fields, the secular part of the homonuclear dipolar interaction is represented as
\begin{align}
    \sum_{i \neq j} \left[ B_{ij} I_{iz} I_{jz}
    + A_{ij} \left(
      I_{i+} I_{j-} + I_{i-} I_{j+}
    \right)
    \right],
\end{align}
whereas that for heteronuclear spins is given by
\begin{align}
        \sum_{i \neq j}  B_{ij} I_{iz} I_{jz}.
\end{align}
Thus, the secular heteronuclear dipolar interactions are special cases of the homonuclear ones, where the coefficient $A_{ij}$ of the flip-flop term may be set to zero if the $i$-th and $j$-th spins belong to different species.
It follows that the Lowe-Gade theory can be applied straightforwardly to systems composed of diferrent spin species.
One practical difficulty in calculating the high-temperature spin diffusion coefficient $D_{0}$ from the atomic coordinates is uncertainty in the actual $^{2}$H substitution sites, for which assumption have to be made.
Nevertheless, the arguments of the polarization-dependence and that of the scaling of $D_{0}$ remain to be legitimate.

\section{Parameters for simulating $^{1}$H-polarization buildup by triplet DNP}
The reasons for using the spin-lattice relaxation rate $1/T_{1}=7.6 \times 10^{-4}$~s$^{-1}$ and the high-temperature spin-diffusion coefficient $D_{0}=4.5 \times 10^{-19}$ m$^{2}$s$^{-1}$ in the simulation presented in Fig.~2 of $^{1}$H polarization buildup by triplet DNP are as follows.

\subsection{Spin-lattice relaxation rate}
\begin{figure} [htbp]
\begin{center}
\includegraphics[width=0.5\linewidth]{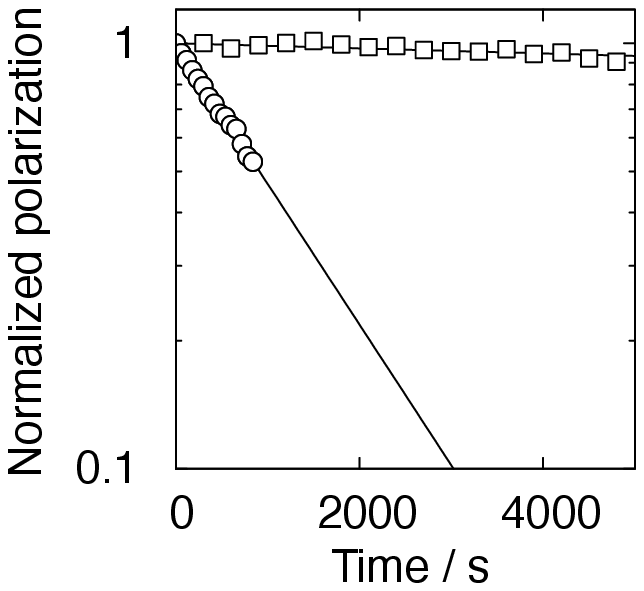}
\caption{Decay of $^{1}$H polarization in a single crystal sample of 99.21 \% deuterated naphthalene doped with pentacene with concentration of $1.46 \times 10^{-4}$ in 0.3187 T and at 100 K, observed after polarization-buildup experiment shown in Fig.~2 under laser irradiation (circles) and without laser irradiation (squares). The pulsed laser irradiation was applied in the same way as was in the triplet DNP experiment, i.e., with a power of 5 mJ/pulse and at a rate of 50 Hz.}
\label{fig:s2}
\end{center}
\end{figure}
In order to estimate the effect of spin-lattice relaxation during the buildup of $^{1}$H polarization by triplet DNP, we observed the decay of the enhanced $^{1}$H polarization toward thermal equilibrium.
Even though the paramagnetic electrons are absent for most of the experimental time in triplet DNP and thus the effect of paramagnetic relaxation is expected to be exceedingly small compared to DNP utilizing permanent paramagnetic electrons, the transient electron paramagnetism ought to have some finite contribution to nuclear relaxation.
Therefore, it is the relaxation under pulsed-laser irradiation which was applied just in the same way as was in the buildup experiment that is of interest.
Fig.~\ref{fig:s2} shows the decay of $^{1}$H polarization observed under pulsed laser irradiation without microwave application, whence we determined the relaxation rate $1/T_{1}$ to be $7.6 \times 10^{-4}$~s$^{-1}$.
When the $^{1}$H spins were just left without laser application, the decaying profile showed much slower behavior, as also shown in Fig.~\ref{fig:s2}.

\subsection{High-temperature spin-diffusion coefficient}
To experimentally determine the spin diffusion coefficient, one can arrange a set of triplet DNP experiments, examining the initial buildup rates for various repetition rates of pulsed laser and microwave applications\cite{Kagawa2009,Miyanishi2021}.
For relatively slow repetition rates such that the rapid-diffusion regime is valid, the initial buildup is proportional to the repetition rate. As increasing the repetition rate and the rapid diffusion limit no longer holds at some point, the initial buildup rate as a function of the repetition rate begins to saturate.
By analyzing the saturation behavior of the initial buildup rate as a function of the repetition rate, the spin diffusion coefficient $D_{0}$ in the high temperature limit can be extracted.

Note that the initial buildup rate is independent of whether the proposed correction to the spin-diffusion coefficient by the factor $1/\sqrt{1-\overline{p}^{2}}$ is adopted or not, because initially $\overline{p} \ll 1$.
For the 99.21 \%-deuterated naphthalene sample, the coefficient $D_{0}$ of spin diffusion among the residual $^{1}$H spins in the high-temperature limit was determined to be $4.5 \times 10^{-19}$ m$^{2}$s$^{-1}$ using this methodology\cite{s-Takeda2009}.

\end{document}